\documentclass[a4paper,12pt]{article}
\usepackage{amssymb}
\usepackage{amsmath}
\usepackage{epsfig}
\usepackage{subfigure}
\usepackage{graphics}

\usepackage{latexsym}
\usepackage{rotating}
\usepackage{titlesec}
\newcommand{\squeezeup}{\vspace{-9.5mm}}

\title{Moving null curves and integrability}

\author{Metin G\"{u}rses \thanks{gurses@fen.bilkent.edu.tr}\\
{\small Department of Mathematics, Faculty of Science}\\
{\small Bilkent University, 06800 Ankara - Turkey}\\
Asl{\i} Pekcan \thanks{Email:aslipekcan@hacettepe.edu.tr} \\
{\small Department of Mathematics, Faculty of Science} \\
{\small Hacettepe University, 06800 Ankara - Turkey}
}

\date{\nonumber}
\begin{document}
\maketitle
\date{\nonumber}
\newtheorem{thm}{Theorem}[section]
\newtheorem{Le}{Lemma}[section]
\newtheorem{defi}{Definition}[section]
\newtheorem{ex}{Example}[section]
\newtheorem{pro}{Proposition}[section]
\baselineskip 17pt

\numberwithin{equation}{section}

\begin{abstract}
We study the null curves and their motion in a $3$-dimensional flat space-time $M_{3}$. We show that when the motion of null curves forms two surfaces in $M_{3}$ the integrability conditions lead to the well-known AKNS hierarchy. In this case we obtain all the geometrical quantities of the surfaces arising from the whole hierarchy but we particulary focus on the surfaces of the MKdV and KdV equations. We obtain  one- and two-soliton surfaces associated to the MKdV equation and show that the Gauss and mean curvatures of these surfaces develop singularities in finite time. We show that the tetrad vectors on the curves satisfy the spin vector equation in the ferromagnetism model of Heisenberg.

\end{abstract}


\section{Introduction}

There are two different ways of studying integrable equations and the surface theory in three dimensional spaces. One of them is to use the Lax equations of certain integrable equations to obtain the associated parametrization of two surfaces in three dimensional Euclidean or Minkowski spaces \cite{sym1}-\cite{Gur1}. The other one is to use the Serret-Frenet equations for curves in three dimensional spaces and defining two surfaces as the traces of the motion of the curves \cite{cramp}-\cite{zhong1}. In this work we shall follow the second approach and study the motion of null curves in Minkowski $3$-space $M_{3}$ and determine surfaces swept by such curves. From the integrability conditions we obtain the well-known Ablowitz-Kaup-Newell-Segur (AKNS) system \cite{akns}. Here we focus on Korteweg-de Vries (KdV) and modified Korteweg-de Vries (MKdV) reductions of the AKNS system and the corresponding two surfaces. In particular, for the MKdV equation we obtain one-soliton and two-soliton surfaces. We obtain the mean and Gauss curvatures of these surfaces and observe that they are singular for the two-soliton surfaces meaning that for some values of the constants these surfaces develop singularities in finite time. We observe  that the tetrad vectors of the curves satisfy the equation of the spin vector $\vec{S}$ in the Heisenberg theory of ferromagnetism and in each case the curvature and the torsion of the curves satisfy certain nonlinear evolution equations.

The layout of the paper is as follows: In Section 2 we give Serret-Frenet equations and motion of the null curves in $n$-dimensional Minkowskian geometry, then in Section 3 we focus on null curves in three dimensional spaces. From the integrability conditions we obtain the recursion operator of the well-known AKNS system and we define surfaces arising from each member of the AKNS hierarchy. In Section 4 we obtain all possible surfaces arising from the hierarchy of AKNS system and derive the mean and Gauss curvatures. In Sections. 5 and 6 we study the surfaces arising from the NLS and MKdV systems. MKdV system has two important reductions KdV and MKdV equations. Surfaces from these equations are very interesting. We present the one- and two-soliton surfaces of the MKdV equation. Finally in Section 7 we considered the spin vector of the Heisenberg ferromagnetism.

\section{Serret-Frenet equations in $M_{3}$}

Two dimensional surfaces in Euclidean $3$-space ${\mathbb R}^3$ and in Minkowski $3$-space $M_{3}$ have been studied for many purposes. One of the main interest in these surfaces is the relation between the integrable evolution equations and these surfaces \cite{cramp}-\cite{zhong1}.

In an $N$-dimensional manifold $M$ a null curve is defined as follows: Let $\vec{e}\,^{a}$ be an $N$-tetrad vectors defined on a curve $\vec{x}(s)$ with tangent vector $\vec{t}=\frac{d \vec{x}}{ds}$. A curve $C$ is called null (or isotropic) if $\vec{t} \cdot \vec{t}=0$ and ${\vec{e}}\,^{a} \cdot {\vec{e}}\,^{b}=g^{ab}$ for all $a, b=1,2, \cdots, N$, where $g^{ab}$ are the components of the inverse of metric tensor $g_{ab}$. In this work we adopt the following metric form of the flat metric in $M$:
\begin{equation}\label{Ndim}
g=\left( \begin{array}{cccccc}
0& 0 & 0 & \cdots&0& 1 \\
0 & 0 &0  & \cdots&1 &0 \\
\cdot & \cdot & \cdot& \cdots & \cdots&0\\
0 & 1 &0 & 0 &\cdots & 0  \\
1 &0 & 0 & 0 &\cdots &0
\end{array} \right).
\end{equation}
At each point of the curve $C$ the tetrad satisfies the Serre-Frenet equations
\begin{equation}
\frac{d \vec{e}\,^{a}}{ds}=\Lambda^{a}\,_{b}\, \vec{e}\,^{b},~~~a=1,2, \cdots, N, \label{sf}
\end{equation}
where $s$ is a parameter of the curve $C$. The components $g_{bc}\, \Lambda^{c}\,_{a}$ must be antisymmetric, i.e.,
\begin{equation}
\Lambda g+g\Lambda^T=0,
\end{equation}
where $g_{ab}$ are the components of the  metric tensor $g^{ab}$. Hence $\Lambda g$ is an antisymmetric matrix in $N$-dimensions \cite{ein}, \cite{carmo}.
 The standard form of the Serret-Frenet equations is given as follows:

First the matrix $\Lambda g$ takes the form
\begin{equation}\label{Ndim}
\Lambda\,g=\left( \begin{array}{cccccc}
0& \kappa & 0 & \cdots&0& 0 \\
-\kappa & 0 &\tau_{1}  & \cdots&0 &0 \\
0 & -\tau_{1} & \cdot& \cdots & 0&0\\
0 & 0 &0 & \cdots &0 & \tau_{N-2}  \\
0 &0 & 0 &\cdots &-\tau_{N-2}&0
\end{array} \right).
\end{equation}
This means that the Serret-Frenet equations are given as
\begin{eqnarray}
&&\frac{d \vec{t}}{ds}= \kappa\, \vec{b}_{N-3},\\
&&\frac{d \vec{n}}{ds}= \tau_{1}\,\vec{b}_{N-4}-\kappa\, \vec{b}_{N-2},\\
&&\frac{d \vec{b}_{1}}{ds}= \tau_{2}\,\vec{b}_{N-5}-\tau_{1}\, \vec{b}_{N-3},\\
&&\cdots \cdots= \cdots \cdots \cdots ,\\
&&\frac{d \vec{b}_{n}}{ds}= \tau_{N-2}\,\vec{n}.
\end{eqnarray}

Motion of the curve is described by $\vec{x}(s,t)$ where $\vec{t}=\frac{\partial \vec{x}(s,t)}{\partial s}$. We let the tetrad frame satisfies the Serret-Frenet equations and equations with respect to $t$ variations
\begin{eqnarray}
&& \frac{\partial{\vec{e}\,^{a}}}{\partial s}=\Lambda^{a}\,_{b}\, \vec{e}\,^{b}, \label{eqn1}\\
&& \frac{\partial{\vec{e}\,^{a}}}{\partial t}=M^{a}\,_{b}\, \vec{e}\,^{b},\label{eqn2}
\end{eqnarray}
where $\Lambda$ matrix is given in (\ref{Ndim}) and $g M$ is an antisymmetric matrix in $N$-dimensions. Integrability of (\ref{eqn1}) and (\ref{eqn2}) gives the zero curvature condition in the theory of integrable systems,
\begin{equation}
\frac{d\Lambda}{dt}-\frac{dM}{ds}=M\Lambda-\Lambda M.
\end{equation}

\section{Moving curves in three dimensions {$N=3$}}

 We consider the Frenet frames on null curves in an $N$-dimensional Minkowski geometry $M_{N}$. First let us consider the case when $N=3$. Let $\vec{e}\,^{a}=(\vec{t}, \vec{n},\vec{b})$ define a Darboux frame with $\vec{e}\,^{a} \cdot \vec{e}\,^{b}=g^{ab}$ where
\begin{equation}
g= \left(\begin{array}{ccc}
             0&0&1\\
             0&1&0\\
             1&0&0\\
             \end{array} \right).
\end{equation}
Then
\begin{equation}
\Lambda= \left(\begin{array}{ccc}
             0&\kappa&0\\
             \tau&0&-\kappa\\
             0&-\tau&0\\
             \end{array} \right),
\end{equation}
where $\kappa$ and $\tau$ are the curvature and torsion functions of the curve, respectively. Then the
Serret-Frenet equations are
\begin{eqnarray}
&&\frac{\partial \vec{t}}{\partial s}= \kappa\, \vec{n}, \\
&&\frac{\partial \vec{n}}{\partial s}= \tau\, \vec{t}- \kappa\, \vec{b}, \\
&&\frac{\partial \vec{b}}{\partial s}= -\tau\, \vec{n}.
\end{eqnarray}
The motion of the curve sweeps the surface $\vec{x}(s,t)$ where the motion of the tetrad at each point is governed by
\begin{eqnarray}
&&\frac{\partial \vec{t}}{\partial t}= f\, \vec{t}+g \, \vec{n},\\
&&\frac{\partial \vec{n}}{\partial t}= h\, \vec{t}-g \, \vec{b},\\
&&\frac{\partial \vec{b}}{\partial t}= -h\, \vec{n}-f \, \vec{b},
\end{eqnarray}
where $f$, $g$, and $h$ are functions of $s$ and $t$ satisfying the integrability conditions
\begin{eqnarray}
&&\frac{\partial \kappa}{\partial t}=\frac{ \partial g}{\partial s}+\kappa\, f, \\
&&\frac{\partial \tau}{\partial t}=\frac{ \partial h}{\partial s}-\tau\, f, \\
&&\frac{\partial f}{\partial s}=\kappa\, h-\tau\,g.
\end{eqnarray}
The last equation can be written as $f=D^{-1}\,(\kappa\,h-\tau\, f)$, where $D^{-1}\,=\int^{s}\, ds$. Then the above integrability conditions turn to be the following evolution equations in matrix form
\begin{equation}
\frac{\partial u}{\partial t}={\mathcal R}\, v
\end{equation}
where $u=(\kappa, \tau)^T$, $v=(g, -h)^T$ and
\begin{equation}\label{recursionAKNS}
{\mathcal R}=\left(\begin{array}{cc}
                 D-\kappa\, D^{-1} \tau & -\kappa D^{-1} \kappa\\
                 \tau D^{-1}\, \tau & -D+\tau\, D^{-1}\, \kappa\\
                 \end{array}\right)
\end{equation}
is the recursion operator of the AKNS system \cite{akns}.

\vspace{0.2cm}
\noindent
Let us choose
\begin{equation}
v={\mathcal R}^{n-1}\,u_{s}
\end{equation}
then $u$ satisfies the AKNS hierarchy
\begin{equation}\label{nlsh}
u_{t}={\mathcal R}^{n}\,u_{s},~~n=0,1,2, \cdots\, .
\end{equation}
For $n=1$ we get the system of NLS equations, for $n=2$ we get the system of MKdV equations, for $n=3$ we get higher order NLS equations etc.
They are respectively given as follows:

\vspace{0.2cm}
\noindent
For $n=1$ we have the system of NLS equations; $g=\kappa_{s}$,~~$h=-\tau_{s}$.
\begin{eqnarray}
&&\kappa_{t}=\kappa_{ss}-\kappa^2\tau,\\
&&\tau_{t}=-\tau_{ss}+\tau^2\kappa.
\end{eqnarray}

\vspace{0.2cm}
\noindent
For $n=2$ we have the system of MKdV equations; $g=\kappa_{ss}-\kappa^2\tau$,~~$h=\tau_{ss}-\tau^2\kappa$.
\begin{eqnarray}
&&\kappa_{t}=\kappa_{sss}-3\kappa\tau\kappa_s,\\
&&\tau_{t}=\tau_{sss}-3\kappa\tau\tau_s.
\end{eqnarray}

\vspace{0.2cm}
\noindent
For $n=3$ we have the system of higher order NLS equations; $g=\kappa_{sss}-3\kappa\tau\kappa_s $, $h=-\tau_{sss}+3\kappa\tau\tau_s$.
\begin{eqnarray}
&&\kappa_{t}=\kappa_{ssss}-3\tau\kappa_s^2-2\kappa\kappa_s\tau_s-4\kappa\tau\kappa_{ss}-\kappa^2\tau_{ss}+\frac{3}{2}\kappa^3\tau^2,\\
&&\tau_{t}=-\tau_{ssss}+3\kappa\tau_s^2+2\tau\kappa_s\tau_s+4\kappa\tau\tau_{ss}+\tau^2\kappa_{ss}-\frac{3}{2}\kappa^2\tau^3.
\end{eqnarray}

\noindent This way we determine infinitely many surfaces for $n \ge 4$.

\noindent Recently, nonlocal reductions of the AKNS system have been also invented \cite{AbMu1}-\cite{AbMu3} and solitonic solutions have been found by Hirota method in \cite{GurPek1}-\cite{gur-pek02}. The surfaces associated to the nonlocal reductions of AKNS hierarchy will be communicated later. Null curves in three dimensions have also been considered in \cite{mus}-\cite{anco}.

\vspace{0.2cm}
\noindent
The AKNS hierarchy has a compatible bi-Hamiltonian structure where ${\mathcal R}=J_{2} \,J_{1}\,^{-1}$. Here  $J_{1}$ and $J_{2}$ are Hamiltonian operators given by
\begin{equation}\label{ham}
J_{2}=\left(\begin{array}{cc}
          -\kappa D^{-1} \kappa&        -D+\kappa\, D^{-1} \tau \\
          -D+\tau\, D^{-1}\, \kappa&       -\tau D^{-1}\, \tau \\
                 \end{array}\right),~~~~
J_{1}=\left(\begin{array}{cc}
          0&  -1 \\
          1&0\\
           \end{array}\right).
\end{equation}
The AKNS hierarchy is given as
\begin{equation}
u_{t}=J_{1}\,\delta \,H_{n+1}=J_{2}\, \delta\, H_{n}, ~~~n=1,2,3, \cdots,
\end{equation}
where $\delta$ is the variational derivative and $H_{n}$'s are the Hamiltonians with
\begin{eqnarray}
&&H_{1}=\frac{1}{2}\, \int_{-\infty}^{\infty}\, \left(\kappa\, \tau_{s}-\tau\, \kappa_{s} \right)\,ds,\\
&&H_{2}=\frac{1}{2}\,\int_{-\infty}^{\infty}\,\left(2 \kappa_{s}\, \tau_{s}+\kappa^2\, \tau^2 \right)\,ds, \\
&&H_{3}=\frac{1}{2}\, \int_{-\infty}^{\infty}\, \left[\tau_{s}\, \kappa_{ss}-\kappa_{s}\, \tau_{ss}+\frac{3}{2} (\kappa \tau^2 \kappa_{s}-\tau \kappa^2 \tau_{s}) \right]\,ds,
\end{eqnarray}
etc. All $H_{n}$'s ($n=1,2, \cdots$) are conserved quantities along the motion of the curves, i.e., $\frac{d H_{n}}{dt}=0$ for all $t$. Here we assume that $\kappa$ and $\tau$ and their $s$ derivatives go to zero as $|s| \to \infty$.

\section{Surfaces swept by null curves}

When $\kappa$ and $\tau$ satisfy equations (\ref{nlsh}) we call them as NLS-system surfaces for $n=1$, MKdV-system surfaces for $n=2$, and higher order NLS-system surfaces for $n \ge 3$. In general, in this sense, they are all integrable surfaces. Now in this section we shall find all geometrical quantities, such as the mean and Gauss curvatures of these surfaces.

Let $S$ be a surface parameterized as $\vec{x}(s,t)$ with the tangent vectors at each point $S$ are given by
\begin{eqnarray}
&&\frac{\partial \vec{x}}{\partial s}=\vec{t},\\
&&\frac{\partial \vec{x}}{\partial t}=A \vec{t}+B \vec{n}+C \vec{b},
\end{eqnarray}
for $A, B, C$ functions of $(s,t)$. Integrability gives
\begin{eqnarray}
&&A_{s}+\tau\, B=f, \\
&&B_{s}+\kappa A-\tau C=g, \\
&&C_{s}-\kappa\, B=0.
\end{eqnarray}
We find that (assuming $\kappa \ne 0$)
\begin{eqnarray}
&&A=\frac{1}{\kappa}\left[\tau\, C-\left(\frac{1}{\kappa}\, C_{s} \right)_{s}+g \right], \\
&&B=\frac{1}{\kappa}\,C_{s},
\end{eqnarray}
where the function $C$ satisfies the following equation
\begin{equation}
\left(\frac{\tau}{\kappa}\, C-\frac{1}{\kappa}\,\left(\frac{1}{\kappa}\, C_{s} \right)_{s}+\frac{g}{\kappa} \right)_{s}+\frac{\tau}{\kappa}\, C_{s}=f.
\end{equation}
Coefficients of the first fundamental form of the surface $S$ are given through the line element given below
\begin{equation}
ds^2=2C ds dt+(2 AC+B^2) dt^2.
\end{equation}
When the function $C$ is not equal to zero (except at some finite number of points) the moving curves form a surface with parameters $s$ and $t$.

Unit normal vector $\vec{N}$ at each point of $S$ is given by $\vec{N}=\epsilon \left(\frac{B}{C}\, \vec{t}-\vec{n} \right)=\epsilon \left(\frac{C_{s}}{\kappa\, C}\, \vec{t}-\vec{n} \right)$, i.e.,
\begin{equation}
\vec{N} \cdot \vec{x}_{s}=0,~~~\vec{N} \cdot \vec{x}_{t}=0, ~~~\vec{N} \cdot \vec{N}=1.
\end{equation}
Here $\epsilon^2=1$. The Weingarten and Gauss equations are respectively given by
\begin{eqnarray}
&&\vec{N}_{,a}=h_{ac}\,g^{ce}\,\vec{x}_{,e}, ~~a=1,2.\\
&&\vec{x}_{,ab}=\Gamma^{c}\,_{ab}\, \vec{x}_{,c}+h_{ab}\, \vec{N}, ~~a,b=1,2.
\end{eqnarray}
where $\Gamma^{a}\,_{bc}$ and $h_{ab}$ are the coefficients of the Christoffel symbol and the second fundamental form, respectively.
We find that
\begin{eqnarray}
&&\vec{N}_{s}=\epsilon\,\left[\left(\frac{C_{s}}{\kappa C}\right)_{s}-\frac{\kappa}{C}\,A-\tau \right]\, \vec{x}_{s}+ \epsilon\,\frac{\kappa}{C}\,\vec{x}_{t},\\
&&\vec{N}_{t}=\epsilon\,  \left( \left(\frac{C_{s}}{\kappa C} \right)_{t}+\frac{C_{s}}{\kappa C}\,f-\frac{g}{C}\,A -h\right)\, \vec{x}_{s}+\epsilon\,\frac{g}{C}\, \vec{x}_{t}.
\end{eqnarray}
Mean curvature $H$ and Gauss curvature $K$ are given by \cite{ein}, \cite{carmo},
\begin{eqnarray}
&&H=\frac{1}{2}\, g^{ab}\,h_{ab}=\frac{\epsilon}{2}\,\left[\left(\frac{C_{s}}{\kappa C}\right)_{s}-\frac{\kappa}{C}\,A-\tau +\frac{g}{C}\right], \\
&&K=\det(g^{-1}\,h)=\frac{g}{C}\,\left[\left(\frac{C_{s}}{\kappa C}\right)_{s}-\tau \right]\,-\frac{\kappa}{C}\,  \left[ \left(\frac{C_{s}}{\kappa C} \right)_{t}+\frac{C_{s}}{\kappa C}\,f-h \right].
\end{eqnarray}
The coefficients $h_{ab}$ of the second fundamental form are
\begin{eqnarray}
&&h_{11}=-\epsilon\, \kappa,~~ h_{12}=h_{21}=-\epsilon \,g, \\
&&h_{22}=-\epsilon \left[Ag+B_{t}-h C -\frac{B}{C}\, (C_{t}-B g-C f) \right].
\end{eqnarray}

\section{NLS-system surfaces}
In this section and the following sections we shall study the specific surfaces for $n=1$ and $n=2$.
For the case $n=1$ we have NLS-system surfaces with $g=\kappa_{s}$, $h=-\tau_{s}$, $f=-\kappa\, \tau$, then
\begin{align}
&\kappa_t=\frac{\partial g}{\partial s}+\kappa f=\kappa_{ss}-\kappa^2\tau,\\
&\tau_t=\frac{\partial h}{\partial s}-\tau f=-\tau_{ss}+\tau^2\kappa.
\end{align}
Here
\begin{equation}
A=\frac{1}{\kappa}\, \left[\tau C-\left(\frac{C}{\kappa}\right)_{s}+\kappa_{s} \right],~~B=\frac{1}{\kappa}\,C_{s}, \label{eqn1A}
\end{equation}
where $C$ satisfies the differential equation
\begin{equation}\label{eqn3}
\left(\frac{\tau}{\kappa}\, C-\frac{1}{\kappa}\,\left(\frac{1}{\kappa}\, C_{s} \right)_{s}+\frac{\kappa_{s}}{\kappa} \right)_{s}+\frac{\tau}{\kappa}\, C_{s}=-\kappa \tau.
\end{equation}
The Gauss and mean curvatures are given as follows:
\begin{eqnarray}
&&H=\frac{\epsilon}{2}\,\left[\left(\frac{C_{s}}{\kappa C}\right)_{s}-\frac{\kappa}{C}\,A-\tau +\frac{\kappa_{s}}{C}\right], \\
&&K=\frac{\kappa_{s}}{C}\,\left[\left(\frac{C_{s}}{\kappa C}\right)_{s}-\tau \right]\,-\frac{\kappa}{C}\,  \left[ \left(\frac{C_{s}}{\kappa C} \right)_{t}-\frac{C_{s}}{C}\, \tau\,+\tau_{s} \right].
\end{eqnarray}

The crucial point here is to solve the function $C$ in (\ref{eqn3}) first and calculate all other functions $A$, $B$, $f$, $g$, and $h$. Then we calculate the mean and Gauss curvatures explicitly in terms of the curvature $\kappa$ and torsion $\tau$ of the null curve. In the case of NLS equation the differential equation (\ref{eqn3}) for $C$ is not so easy to solve. For this reason we shall consider the cases for KdV and MKdV equations in the next sections.

\section{MKdV-system surfaces}
For $n=2$, we have $g=\kappa_{ss}-\kappa^2\tau$, $h=\tau_{ss}-\tau^2\kappa$ yielding
the MKdV system
\begin{eqnarray}
&&\kappa_{t}=\kappa_{sss}-3\kappa\tau\kappa_s,\\
&&\tau_{t}=\tau_{sss}-3\kappa\tau\tau_s.
\end{eqnarray}
Since
\begin{equation}
\frac{\partial \kappa}{\partial t}=\frac{\partial g}{\partial s}+\kappa f=\kappa_{sss}-3\kappa\tau\kappa_s,
\end{equation}
we have
\begin{equation}
f=\kappa\tau_s-\tau\kappa_s.
\end{equation}
We have two important reductions of these surfaces.

\subsection{KdV surfaces}
The KdV equation corresponds to the choice $\tau=\tau_{0}=$ constant in MKdV system, i.e.,
\begin{equation}\label{KdVeq}
\kappa_{t}=\kappa_{sss}-3 \tau_{0}\, \kappa\,\kappa_s.
\end{equation}
In this case we have
\begin{equation}
g=\kappa_{ss}-\tau_0\kappa^2,\quad h=-\tau_0^2\kappa,\quad f=-\tau_0\kappa_s.
\end{equation}
The functions $A$ and $B$ become
\begin{eqnarray}
&&A=\frac{1}{\kappa}\left[\tau_0\, C-\left(\frac{1}{\kappa}\, C_{s} \right)_{s}+\kappa_{ss}-\tau_0\kappa^2 \right], \\
&&B=\frac{1}{\kappa}\,C_{s},
\end{eqnarray}
where $C$ satisfies
\begin{equation}
\left(\frac{\tau_0}{\kappa}\, C-\frac{1}{\kappa}\,\left(\frac{1}{\kappa}\, C_{s} \right)_{s}+\frac{1}{\kappa}(\kappa_{ss}-\tau_0\kappa^2) \right)_{s}+\frac{\tau_0}{\kappa}\, C_{s}=-\tau_0\kappa_s.
\end{equation}
Here mean $H$ and Gauss $K$ curvatures are
\begin{eqnarray}
&&H=\frac{\epsilon}{2}\,\left[\left(\frac{C_{s}}{\kappa C}\right)_{s}-\frac{\kappa}{C}\,A-\tau_0 +\frac{1}{C}(\kappa_{ss}-\kappa^2\tau_0)\right], \\
&&K=\frac{1}{C}(\kappa_{ss}-\kappa^2\tau_0)\,\left[\left(\frac{C_{s}}{\kappa C}\right)_{s}-\tau_0 \right]\,-\frac{\kappa}{C}\,  \left[ \left(\frac{C_{s}}{\kappa C} \right)_{t}-\frac{C_{s}}{\kappa C}\tau_0\kappa_s\,+\tau_0^2\kappa \right].
\end{eqnarray}
\medskip

\noindent Let $\tau_0=2$ and
\begin{equation}\label{kappaKdV}
\kappa=-2\partial_s^2(\ln F).
\end{equation}
Then the equation (\ref{KdVeq}) turns to be
\begin{align}
\frac{2}{F^3}(F^2F_{5s}-5FF_sF_{4s}+2FF_{ss}F_{sss}+8FF_{sss}F_s^2 -6F_sF_{ss}^2-&F^2F_{sst}+FF_tF_{ss}\nonumber\\
&+2FF_sF_{s}-2F_s^2F_t)=0,
\end{align}
which is equivalent to
\begin{equation}\displaystyle
\Big(\frac{(D_s^4-D_sD_t)\{F\cdot F\}}{F^2}\Big)_s=0.
\end{equation}
Here $D_j$ is a special differential operator called Hirota $D$-operator \cite{hir1}-\cite{hir3} given by
\begin{equation}
D_t^nD_x^m\{F\cdot G\}=\Big(\frac{\partial}{\partial t}-\frac{\partial}{\partial t'}\Big)^n\Big(\frac{\partial}{\partial x}-\frac{\partial}{\partial x'}\Big)^m\,F(x,t)G(x',t')|_{x'=x,t'=t}
\end{equation}
for $m$ and $n$ positive integers, and $F, G$ are differentiable functions.

\noindent Under (\ref{kappaKdV}) we have
\begin{align}
&g=\frac{2}{F^4}[2F_s^4-F^2F_{ss}^2-4FF_s^2F_{ss}+4F^2F_sF_{sss}-F^3F_{4s}],\\
&h=\frac{4}{F^2}[FF_{ss}-F_s^2],\\
&f=\frac{4}{F^3}[2F_s^3-3FF_sF_{ss}+F^2F_{sss}],\\
&A=-\frac{F^2}{2(FF_{ss}-F_s^2)}\Big[2C+\Big(\frac{F^2C_s}{FF_{ss}-F_s^2}\Big)_s+\frac{2}{F^4}(2F_s^4-F^2F_{ss}^2-4FF_s^2F_{ss}\nonumber\\
&\hspace{9.8cm}+4F^2F_sF_{sss}-F^3F_{4s})\Big],\\
&B=-\frac{F^2C_s}{2(FF_{ss}-F_s^2)},
\end{align}
where $C$ is satisfying
\begin{align}
&\Big[-\frac{F^2C}{(FF_{ss}-F_s^2)}-\frac{F^2}{4(FF_{ss}-F_s^2)}\Big( \frac{F^2C_s}{FF_{ss}-F_s^2}\Big)_s
-\frac{1}{F^2(FF_{ss}-F_s^2)}(2F_s^4-F^2F_{ss}^2\nonumber\\
&-4FF_s^2F_{ss}+4F^2F_sF_{sss}-F^3F_{4s})\Big]_s
-\frac{F^2C_s}{FF_{ss}-F_s^2}=\frac{4}{F^3}[2F_s^3-3FF_sF_{ss}+F^2F_{sss}].\nonumber\\
\end{align}
Mean and Gauss curvatures become
\begin{align}
H=&\frac{\epsilon}{2}\Big[-\frac{1}{2}\Big(\frac{F^2C_s}{C(FF_{ss}-F_s^2)}\Big)_s+\frac{2A}{C}(FF_{ss}-F_s^2)-2\nonumber\\
&\hspace{5cm}+\frac{2}{CF^4} [2F_s^4-F^2F_{ss}^2-4FF_s^2F_{ss}+4F^2F_sF_{sss}-F^3F_{4s} ] \Big],\\
K=&\frac{2}{CF^4} [2F_s^4-F^2F_{ss}^2-4FF_s^2F_{ss}+4F^2F_sF_{sss}-F^3F_{4s} ]\Big[-\frac{1}{2}\Big(\frac{F^2C_s}{C(FF_{ss}-F_s^2)}\Big)_s-2\Big]\nonumber\\
&+\frac{2}{F^2C}(FF_{ss}-F_s^2)\Big[-\frac{1}{2}\Big(\frac{F^2C_s}{C(FF_{ss}-F_s^2)}\Big)_t+\frac{2C_s}{CF}(3FF_sF_{ss}-2F_s^3-F^2F_{sss})\nonumber\\
&\hspace{12cm}-\frac{8}{F^2}(FF_{ss}-F_s^2)\Big].
\end{align}

\subsection{MKdV surfaces}
MKdV surfaces correspond to $\tau=k_{0}\, \kappa$ where $k_{0}$ is an arbitrary constant, i.e.,
\begin{equation}\label{mKdVeq}
\kappa_t-\kappa_{sss}+3k_0\kappa^2\kappa_s=0.
\end{equation}
We have $g=\kappa_{ss}-k_{0} \kappa^3$, $h=k_{0}\,g$, $f=0$, and $C=\frac{1}{2}\, \kappa^2+\alpha_{0}(t)$ giving
\begin{equation}
A=k_0(-\frac{1}{2}\kappa^2+\alpha_0(t)),\quad B=\kappa_{s}.
\end{equation}
Then we can easily calculate all other geometrical quantities
\begin{eqnarray}
&&H=\frac{\epsilon}{(\kappa^2+2\alpha_0(t))^2}[2\kappa_{ss}(\kappa^2+2\alpha_0(t))-2\kappa\kappa_s^2-k_0\kappa(\kappa^2+2\alpha_0(t))^2],\\
&&K=\frac{4}{(\kappa^2+2\alpha_0(t))^3}[(\kappa^2+2\alpha_0(t))\kappa_{ss}^2-2\kappa\kappa_s^2\kappa_{ss}-k_0\kappa^3(\kappa^2+2\alpha_0(t))\kappa_{ss}\nonumber\\
&&\hspace{3.5cm}-\kappa(\kappa^2+2\alpha_0(t))\kappa_{st}+2k_0\kappa^4\kappa_s^2+2\kappa^2\kappa_s\kappa_t+2\kappa\kappa_s\frac{d \alpha_{0}(t)}{dt}].
\end{eqnarray}

\vspace{0.2cm}
\noindent
We now present some MKdV surfaces obtained from the soliton solutions of the MKdV equation.

\vspace{0.2cm}
\noindent
{\bf One-soliton surfaces}: \,Let $k_0=-8$ and $\kappa=\frac{p_sr-pr_s}{p^2+r^2}$. The equation (\ref{mKdVeq}) can be written in Hirota bilinear form as
\begin{align}
&(D_s^3-D_t)\{p\cdot r\}=0,\label{MKdVH-a}\\
&D_s^2\{p\cdot p+r\cdot r\}=0.\label{MKdVH-b}
\end{align}
One-soliton solution of (\ref{mKdVeq}) is obtained by $p=e^{\theta_1}, r=1$, where $\theta_1=k_1s+\omega_1t+\delta_1$ for $k_1, \omega_1, \delta_1$ constants.
The Hirota bilinear form (\ref{MKdVH-a}) and (\ref{MKdVH-b}) gives the dispersion relation $\omega_1=k_1^3$. Therefore one-soliton solution becomes
\begin{equation}\displaystyle
\kappa=\frac{k_1e^{\theta_1}}{1+e^{2\theta_1}},\quad \theta_1=k_1s+k_1^3t+\delta_1.
\end{equation}
Under the above solution we have
\begin{align}\displaystyle
&A=\frac{4k_1^2e^{2\theta_1}}{(1+e^{2\theta_1})}-8\alpha_0(t),\\
&B=\frac{k_1^2e^{\theta_1}(1-e^{2\theta_1})}{(1+e^{2\theta_1})^2},\\
&C=\frac{[(k_1^2+4\alpha_0(t))+2\alpha_0(t)(1+e^{4\theta_1})]}{2(1+e^{2\theta_1})^2},
\end{align}
and mean and Gauss curvatures become
\begin{align}\displaystyle
&H=\frac{4\epsilon k_1\alpha_0(t)(k_1^2+8\alpha_0(t))e^{\theta_1}(1+e^{2\theta_1})^3}{[(k_1^2+4\alpha_0(t))e^{2\theta_1}+2\alpha_0(t)(1+e^{4\theta_1})]^2}, \label{gaus1}\\
&K=\frac{8k_1^3e^{2\theta_1}\frac{d \alpha_{0}(t)}{dt}[1+2e^{2\theta_1}-2e^{6\theta_1}-e^{8\theta_1}] }{[(k_1^2+4\alpha_0(t))e^{2\theta_1}+2\alpha_0(t)(1+e^{4\theta_1})]^3}. \label{gaus2}
\end{align}
Note that if $\alpha_0(t)=0$ we have $H=K=0$. For $\alpha_0(t)=\mathrm{constant}\neq 0$ we have $K=0$ but $H\neq 0$. Consider the following example.

\noindent \textbf{Example 1.} Using one-soliton solution for $\kappa$ with $\alpha_0(t)=\mathrm{constant}$, say $\alpha_0(t)=1$, and taking $\epsilon=1, k_1=\frac{3}{2}, \delta_1=0$ give $K=0$ and
\begin{equation}
H=\frac{123}{2}\frac{e^{\frac{3}{2}s+\frac{27}{8}t}(1+e^{3s+\frac{27}{4}t})^3}{(\frac{25}{4}e^{3s+\frac{27}{4}t}+2+2e^{6s+\frac{27}{2}t} )^2}.
\end{equation}
The graph of the above mean curvature is given in Figure 1.
\begin{center}
\begin{figure}[h]
\centering
\begin{minipage}[t]{1\linewidth}
\centering
\includegraphics[angle=0,scale=.33]{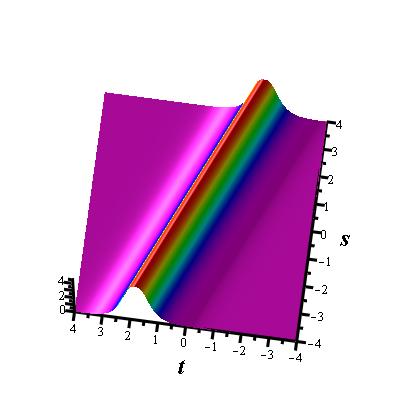}
\caption{The graph of the mean curvature $H$ of MKdV surface
with the parameters $\alpha_0(t)=1$, $\epsilon=1, k_1=\frac{3}{2}, \delta_1=0$.}
\end{minipage}%
\end{figure}
\end{center}
\squeezeup
It is clear that the mean curvature has no singularities for all $s$ and $t$.

\vspace{0.2cm}
\noindent
{\bf Two-soliton surfaces}: Two-soliton solution of the MKdV equation (\ref{mKdVeq}) can also be obtained by the help of Hirota method. Take $p=e^{\theta_1}+e^{\theta_2}$ and $r=1+A_{12}e^{\theta_1+\theta_2}$ in the Hirota bilinear form (\ref{MKdVH-a}) and (\ref{MKdVH-b}). Here $\theta_j=k_js+\omega_jt+\delta_j$, $j=1, 2$. We get
$\omega_j=k_j^3$, $j=1,2$, and $A_{12}=-\frac{(k_1-k_2)^2}{(k_1+k_2)^2}$. Therefore two-soliton solution of (\ref{mKdVeq}) is
\begin{equation}\displaystyle
\kappa=\frac{k_1e^{\theta_1}+k_2e^{\theta_2}-A_{12}e^{\theta_1+\theta_2}(k_1e^{\theta_2}+k_2e^{\theta_1}) }{1+e^{2\theta_1}+e^{2\theta_2}+2e^{\theta_1+\theta_2}(1+A_{12})+A_{12}^2e^{2\theta_1+2\theta_2} }. \label{kap}
\end{equation}
If we use two-soliton solution for $\kappa$ the expressions for $A,B,C, H,$ and $K$ become lengthy. But here we note that even if $\alpha_0(t)=0$, we have $H\neq 0$ and $K\neq 0$. In the following example we present graphs of $H$ and $K$ for particular choice of solution parameters with $\alpha_0(t)=0$.

\noindent \textbf{Example 2.} Using two-soliton solution for $\kappa$ with $\alpha_0(t)=0$, and taking $\epsilon=1, k_1=\frac{1}{4}, k_2=2, \delta_1=\delta_2=0$ give the graphs of $H$ and $K$ in Figure 2 and Figure 3, respectively.
\begin{center}
\begin{figure}[h]
\centering
\begin{minipage}[t]{0.41\linewidth}
\centering
\includegraphics[angle=0,scale=.33]{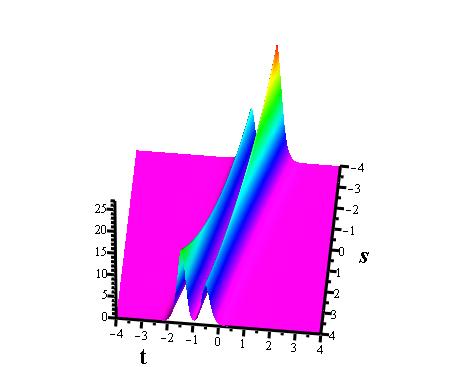}
\caption{The graph of the mean curvature $H$ of MKdV surface with the parameters $\alpha_0(t)=0$, $\epsilon=1, k_1=\frac{1}{4}, k_2=2, \delta_1=\delta_2=0$.}
\end{minipage}%
\hfill
\begin{minipage}[t]{0.41\linewidth}
\centering
\includegraphics[angle=0,scale=.33]{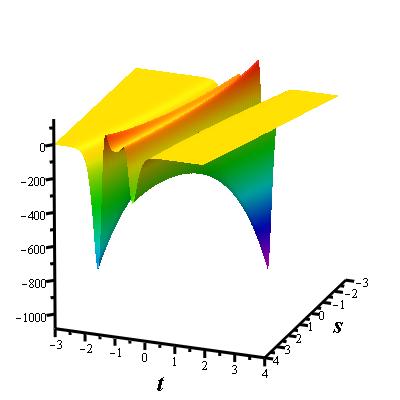}
\caption{The graph of the Gauss curvature $K$ of MKdV surface with the parameters $\alpha_0(t)=0$, $k_1=\frac{1}{4}, k_2=2, \delta_1=\delta_2=0$.}
\end{minipage}%
\end{figure}
\end{center}
\squeezeup
\noindent
We observe that both the mean and Gauss curvatures develop singularities for finite values of the parameter $t$. Such singular behaviour comes from the singularities of the curvature $\kappa$ given in (\ref{kap}). These singularities arise for particular values of the function $\alpha_{0}(t)$ in (\ref{gaus1}) and (\ref{gaus2}) and on $k_{1}$ and $k_{2}$ in two-soliton solutions.

\section{Spin vector of the Heisenberg model}
When one studies motion of the tetrad on a curve it is custom to express the spin vector $\vec{S}$ in the Heisenberg's ferromagnetism model in terms of the tetrad vectors. In this model $\vec{S}$ satisfies the vector differential equation  (see for instance \cite{ding1}, \cite{lak1} )
\begin{equation}
\vec{S}_{t}=\vec{S} \times \vec{S}_{ss}
\end{equation}
For this purpose we need the vector products of the tetrad vectors

\begin{eqnarray}
&&\vec{t} \times \vec{n}=\vec{t}, \\
&&\vec{t} \times \vec{b}=-\vec{n}, \\
&&\vec{n} \times \vec{b}=\vec{b}.
\end{eqnarray}
Using the above vector products we have the following cases:

\vspace{0.3cm}
\noindent
{\bf 1}. Let $\vec{S}=\vec{t}$, then $f=\kappa_{s}, ~~B=0, ~~g=\kappa^2$ and $h \kappa=\tau g+f_s$. In this case the pair $(\kappa, \tau)$ satisfies
the following evolution equations:
\begin{eqnarray}
&&\kappa_{t}=3 \kappa \kappa_{s}, \\
&&\tau_{t}=\left(\frac{\kappa_{ss}}{\kappa}\right)_{s}+\kappa \tau_{s}.
\end{eqnarray}
When $\kappa$ is a nonzero constant then the torsion of the curve
satisfies the linear equation $\tau_{t}=\kappa\, \tau_{s}$.

\vspace{0.3cm}
\noindent
{\bf 2}. Let $\vec{S}=\vec{n}$, then $h=-\tau_s$, $g=\kappa_s$, and  $f_s=h\kappa-g\tau$.
Here the pair $(\kappa, \tau)$ satisfies
\begin{eqnarray}
&&\kappa_{t}=\kappa_{ss}-\kappa^2\tau+\kappa \alpha_1(t), \\
&&\tau_{t}=-\tau_{ss}+\kappa\tau^2-\tau\alpha_1(t).
\end{eqnarray}
If $\alpha_1(t)=0$, we have the NLS system.\\

\vspace{0.3cm}
\noindent
{\bf 3}. Let $\vec{S}=\vec{b}$, then $h=\tau^2$, $f=-\tau_s$, and $g\tau=h\kappa-f_s$
In this case the functions $\kappa$ and $\tau$ satisfy
\begin{eqnarray}
&&\kappa_{t}=\Big(\frac{\tau_{ss}}{\tau}\Big)_s+\tau\kappa_s, \\
&&\tau_{t}=3\tau\tau_s.
\end{eqnarray}

\section{Concluding Remarks}

In this work we focused on moving null curves in a three dimensional Minkowski space $M_{3}$ and considered the case they form two surfaces in $M_{3}$.
We showed that the integrability conditions lead to the AKNS hierarchy. This means that we obtain infinitely many surfaces corresponding to each member of the hierarchy. As examples we studied the surfaces arising from the KdV and MKdV equations. In particular we obtained one- and two-soliton surfaces of the MKdV equation. We observed that the tetrad vectors satisfy the spin equations in Heisenberg model of ferromagnetism. In all the possible cases the curvature and the torsion of the curves satisfy certain nonlinear partial differential equations. For the two cases the spin vectors are null but in the case where the spin vector is the normal vector the spin vector is a spacelike vector and curvature and the torsion pair satisfies the NLS.

\section{Acknowledgment}
  This work is partially supported by the Scientific
and Technological Research Council of Turkey (T\"{U}B\.{I}TAK).\\


\begin{thebibliography}{}

\bibitem{sym1} A. Sym, {\it Soliton surfaces}, Lett.Nuovo Cimento Soc. Ital.Fis. {\bf 33}, 394 (1982).

\bibitem{sym2} A. Sym, {\it Soliton surfaces II}, Lett.Nuovo Cimento Soc. Ital.Fis. {\bf 36}, 307 (1983).

\bibitem{sym3} A. Sym, {\it Soliton surfaces III: Solvable nonlinearities with trivial geometry}, Lett. Nuovo Cimento Soc. Ital. Fis.{\bf 39}, 193 (1984).

\bibitem{fok1} A.S. Fokas, I.M. Gel'fand, {\it Surfaces on Lie groups, on Lie algebras and their integrability}, Commun. Math. Phys. {\bf 177}, 203 (1996).

\bibitem{fok2} A.S. Fokas, I.M. Gel'fand, F. Finkel, and Q.M. Liu, {\it A formula for constructing infinitely many surfaces on Lie algebras and integrable equations}, Sel. Math. (New Ser.) \textbf{6}, 347--375 (2000).

\bibitem{cey} \"{O}. Ceyhan, A.S. Fokas, and M. G{\" u}rses, {\it Deformations of surfaces associated with integrable Gauss-Meinardi-Codazzi equations}, J. Math. Phys.  {\bf 41}, 2251--2270 (2000).


\bibitem{Gur1} Metin G{\" u}rses, S\"{u}leyman Tek, {\it Integrable Curves and Surfaces}, A Talk in XVI.th Conference on "Integrability, Geometry and Quantization,
Eds. I. Mladenov, A. Ludu, and A. Yoshioda, June 5-11, 2015, Varna.

\bibitem{cramp} M. Crampin, F.A.E. Pirani, and D.C. Robinson, {\it The soliton connection}, Lett. Math. Phys. {\bf 2}, 15--19 (1977).

\bibitem{gur} M. G\"{u}rses, Y. Nutku, {\it New nonlinear evolution equations from surface theory}, J. Math. Phys. {\bf 22}, 1393 (1981).

\bibitem{konop} B.G. Konopelchenko, {\it Induced surfaces and their integrable dynamics}, Stud. Appl. Math. {\bf 96}(1), 9 (1996).

\bibitem{has} H. Hasimoto, {\it A soliton on a vortex filament}, J. Fluid Mech. {\bf 51}(3), 477--485 (1972).

\bibitem{lamb1} G.L. Lamb, {\it Solitons and the motion of helical curves}, Phys. Rev. Lett. {\bf 37}, 235 (1976).

\bibitem{lamb2} G.L. Lamb, {\it Solitons on moving space curves}, J. Math. Phys, {\bf 18} 1654--1661 (1977).

\bibitem{wadat} K. Nakayama, H. Segur, and M. Wadati, {\it Integrability and the motion of curves}, Phys. Rev. Lett. {\bf 69}, 2603 (1992).

\bibitem{laks} M. Lakshmanan, {\it Rigid body motions, space curves, prolongation structures, fiber bundles, and solitons}, J. Math. Phys. {\bf 20}, 1667--1672 (1978).

\bibitem{pet}  R.E. Goldstein, D.M. Petrich, {\it The Korteweg–de Vries hierarchy as dynamics of closed curves in the plane}, Phys. Rev. Lett. {\bf 67}, 3203 (1991).

\bibitem{cies} J. Cie\'{s}li\'{n}ski, P.K.H. Gragert, and A. Sym, {\it Exact solution to localized-induction-approximation equation modeling smoke ring motion}, Phys. Rev. Lett. {\bf 57}, 1507 (1986).

\bibitem{lang} J. Langer, R. Perline, {\it Poisson geometry of the filament equation}, J. Nonlinear Sci. {\bf 1}, 71--93 (1991).


\bibitem{gur0} M. G\"{u}rses, {\it Motion of curves on two-dimensional surfaces and soliton equations}, Phys. Lett. A {\bf 241}, 329--334 (1998).

\bibitem{ding1} Q. Ding, J. Inoguchi, {\it Schr\"{o}dinger flows, binomial motion for curves and the second AKNS-hierarchies}, Chaos Solitons Fractals {\bf 21}, 669--677 (2004).

\bibitem{zhong1} S. Zhong, {\it A motion of complex curves in ${\mathbb C}^3$ and nonlocal nonlinear Schr\"{o}dinger equation}, J. Nonlinear Sci. Appl. {\bf 12}, 75--85 (2019).

\bibitem{akns} M.J. Ablowitz, D.J. Kaup, A.C. Newell, and H. Segur, {\it The inverse scattering transform-Fourier analysis for nonlinear problems}, Stud. Appl. Math. \textbf{53} (4),  249--315, (1974).


\bibitem{ein} L.P. Eisenhart, {\it A Treatise on the Differential Geometry of Curves and Surfaces} (Ginn. Boston, 1909, reprinted Dover, New York 1960).

\bibitem{carmo} M.P. do Carmo, {\it Differential Geometry of Curves and Surfaces} (Prentice-Hall, Englewood Cliffs, NJ).

\bibitem{AbMu1} M.J. Ablowitz, Z.H. Musslimani, {\it Integrable nonlocal nonlinear Schr\"{o}dinger equation}, Phys. Rev. Lett. \textbf{110}, 064105, (2013).

\bibitem{AbMu2} M.J. Ablowitz, Z.H. Musslimani, {\it Inverse scattering transform for the integrable nonlocal nonlinear Schr\"{o}dinger equation}, Nonlinearity \textbf{29} 915--946, (2016).

\bibitem{AbMu3} M.J. Ablowitz, Z.H. Musslimani,  {\it Integrable nonlocal nonlinear equations}, Stud. Appl. Math. \textbf{139} (1), 7--59, (2016).


\bibitem{GurPek1}  M. G\"{u}rses, A. Pekcan, {\it Nonlocal nonlinear Schr\"{o}dinger equations and their soliton solutions}, J. Math. Phys. \textbf{59}, 051501, (2018).

\bibitem{GurPek3} M. G\"{u}rses, A. Pekcan, {\it Integrable Nonlocal Reductions}, "Symmetries, Differential Equations and Applications SDEA-III, Istanbul, Turkey, August 2017", Editors: V.G. Kac, P.J. Olver, P. Winternitz, and T. Ozer, Springer Proceedings in Mathematics and Statistics, \textbf{266}, (2018).

\bibitem{GurPek2}  M. G\"{u}rses, A. Pekcan, {\it Nonlocal nonlinear modified KdV equations and their soliton solutions},
Comm. Non. Sci. Numer. Simulat. \textbf{67}, 427--448, (2019).

\bibitem{GurPek4} M. G{\" u}rses, A. Pekcan, {\it $(2+1)$-dimensional local and nonlocal reductions of the
negative AKNS system: Soliton solutions}, Commun. Nonlinear Sci. Numer. Simulat. \textbf{71}, 161--173, (2019).

\bibitem{gur-pek02}  M. G{\" u}rses, A. Pekcan, {\it $2+1$-dimensional AKNS($-N$) systems II}, Commun. Nonlinear Sci. Numer. Simulat. {\bf 92}, 105736, (2021).


\bibitem{mus} E. Musso, L. Nicolodi, {\it Hamiltonian flows on null curves}, Nonlinearity \textbf{23}, 2117--2129 (2010).

\bibitem{gaber} S. Gaber and A.A. Elaiw, {\it Inextensible flows of null Cartan curves in Minkowski space ${\mathbb R}^{2,1}$}, Universe {\bf 9}, 125 (2023).

\bibitem{anco} Z.K. Y{\" u}zba\c{s}{\i}, S.C. Anco, {\it Elastic null curve flows, nonlinear C-integrable systems and geometric realization of Cole-Hopf transformations}, J. Nonlinear Math. Phys. {\bf 27}, 357 (2021).


\bibitem{hir1} R. Hirota, {\it The Direct Method in Soliton Theory}, Cambridge University Press, Cambridge, (2004).

\bibitem{hir2} R. Hirota, {\it Exact solution of the Korteweg-de Vries equation for multiple
collisions of solitons}, Phys. Rev. Lett. \textbf{27} (1971) 1192.

\bibitem{hir3} R. Hirota, {\it Exact solution of the modified Korteweg-de Vries equation for
multiple collisions of solitons}, J. Phys. Soc. Japan \textbf{33} (1972) 1456.


\bibitem{lak1} M. Lakshmanan, {\it Continuum spin system as an exactly solvable dynamical system}, Phys. Lett. A {\bf 61}, 53 (1979).








\end{thebibliography}
\end{document}